\documentclass[12pt,preprint]{aastex}

\def\Eiso{$E_{\rm iso}$ }
\def\tsutsui{$E_p$--$T_L$--$L_p$ }
\def\yonetoku{$E_p$--$L_p$ }
\def\amati{$E_p$--$E_{\rm iso}$ }

\slugcomment{Not to appear in Nonlearned J., 45.}

\shorttitle{Gold data set and the \tsutsui relation of GRBs}
\shortauthors{Tsutsui et al.}

\begin{document}

\title{Intrisic Dispersion of   Correlations among $E_p, L_p$, and \Eiso of Gamma Ray Bursts
 depends on the quality of Data Set}


\author{R. Tsutsui and T. Nakamura}
\affil{Department of Physics, Kyoto University,
Kyoto 606-8502, Japan}
\email{tsutsui@tap.scphys.kyoto-u.ac.jp}

\author{D. Yonetoku and T. Murakami}
\affil{Department of Physics, Kanazawa University,
Kakuma, Kanazawa, Ishikawa 920-1192, Japan}

\author{K.Takahashi}
\affil{Department of Physics and Astrophysics,
Nagoya University, Fro-cho, Chikusa-ku, Nagoya, 464-8602, Japan
}



\begin{abstract}

We reconsider correlations among the spectral peak energy ($E_p$), 
1-second peak luminosity ($L_p$) and isotropic energy (\Eiso),
using the database constructed by \citet{yonetoku10}
which consists of 109 Gamma-Ray Bursts (GRBs) whose redshifts are
known and $E_p$, $L_p$ and \Eiso are well determined.
We divide the events into two groups by their data quality.
One (gold data set) consists of GRBs with peak energies
determined by the Band model with four free parameters.
On the other hand, GRBs in the other group (bronze data set)
have relatively poor energy spectra so that their peak energies
were determined by the Band model with fixed spectral index
(i.e. three free parameters) or by the Cut-off power law (CPL) model
with three free parameters. Using only the gold data set we found
the intrinsic dispersion in $\log L_p$ ($=\sigma_{\rm int}$) is 0.13
and 0.22 for \tsutsui correlation ($T_L \equiv E_{\rm iso}/L_p$) and
\yonetoku correlation, respectively. We also find that GRBs
in the bronze data set have systematically larger $E_p$ than
expected by the correlations constructed with the gold data set.
This means that the intrinsic dispersion of correlations among
$E_p$, $L_p$, and \Eiso of GRBs depends on the quality of data set.
At present, using \tsutsui correlation with gold data set,
we would be able to determine the luminosity distance with
$\sim 16\%$ error, which might be useful to determine
the nature of the dark energy at high redshift $z > 3$.\\
\end{abstract}





\section{Introduction}
\label{sec:introduction}

Discoveries of empirical correlations of gamma-ray bursts (GRBs) 
raised many researches on early universe using GRBs. 
One of the most well studied correlations is the one between
the spectral peak energy ($E_p$) and isotropic equivalent energy (\Eiso)
called \amati correlation \citep{amati02,sakamoto04,lamb04,amati06,amati09}.
\citet{yonetoku04} found a similar but tighter correlation between 
$E_p$ and 1-second peak luminosity called the \yonetoku correlation.
These correlations are  tight  but they have large dispersions 
such as $\sigma_{int}=0.33$ in $\log L_p$ and 
$\sigma_{int}=0.37$ in $\log$\Eiso  which can not be explained 
as statistical errors of $E_p$ , \Eiso  and $L_p$ \citep{yonetoku10}.
\citet{ggl04} found that $E_p$ tightly correlates with 
the collimation-corrected gamma-ray energy ($E_{\gamma}$).
\citet{firmani06} proposed that adding the high signal 
time scale ($T_{0.45}$) to the \yonetoku relation reduces the dispersion
of the correlation. This correlation is defined by using only prompt
emission properties like \amati, \yonetoku  correlations
so that it seems to be promising tools to constrain the cosmological
parameters. However, this correlation is not confirmed by later
studies \citep{rossi08, collazzi08}. More recently,
\citet{tsutsui09} found that  the luminosity time
($T_L=E_{\rm iso} / L_p$) also improves both the \amati and
\yonetoku correlations. 

These correlations were used to investigate the star formation history
\citep{yonetoku04}, the reionaization epoch \citep{murakami05},
and the cosmological expansion history of the early universe
\citep{takahashi,oguri,ghirlanda06,schaefer07, kodama08,
liang08,cardone09, tsutsui09}.

However, in spite of high correlation coefficients, there have been
many cautions to use these empirical correlations for cosmology
\citep{np04,bp05,butler,sn09}. To establish these correlations
in GRBs prompt emissions as tools to determine cosmological parameters,
we must investigate  the origins of  systematic errors and the way
to remove them. We note that there are many factors to cause systematic
errors besides intrinsic dispersions of their prompt emissions.
For  example the sensitivity of the detectors , the evolution effects
of GRBs, the confusion with other sources, the lack of unknown
parameters like the jet opening angle $\theta_{jet}$, etc.
All of these effects might arise the additional systematic errors
over the  intrinsic dispersions  of GRBs.

Possible selection effects on these correlations are studied by
many authors with contrasting results
\citep{butler, ghirlanda08, nava08, sn09, amati09, yonetoku10}.
However previous studies did not consider the difference of 
spectral models to determine $E_{p}$ well. As shown in \cite{kaneko06},
it often happens that high energy power-law index $\beta$ for
the Band model with four free parameters can not be determined
by the data so that the cutoff power-law (CPL) model with three
free parameters is used to fit the data. CPL model might be good
if the peak energy is close to the high energy end of the detector
band width. One can not use Band model but CPL model if the event
is so dim that the number of high energy photons
is very small. Importantly, simulations in \cite{kaneko06}
showed that, if the signal-to-noise ratio is relatively low
, a true spectrum with the shape of the Band model  can be fitted
by CPL  model with $E^{\rm obs}_p$ which is larger
than the true value of $E^{\rm obs}_p$  up to $\sim 100~{\rm keV}$.
Therefore CPL model might overestimate $E_p$.
While , if we fit a true CPL  spectrum  
by the Band model, the estimated value of $E_{p}^{obs}$ is almost equal to 
the true value since the large value of $-\beta$ looks like an exponential function.  
In reality \citet{sn09} found that $E_{p}$ estimated using the CPL model by \citet{kaneko06}
are systematically harder than $E_{p}$ estimated using the Band model by \citet{yonetoku04}.
Although the systematic difference between the peak energies 
fitted by the Band model and the ones fitted by the CPL model are 
reported, how this difference affect the spectral-brightness correlations 
of GRBs has been hardly studied so that we shall study this problem in this paper.

The purpose of this letter is to investigate the effect of uncertainty
in using different spectral models which determine $E_p$
on the \amati , \yonetoku and \tsutsui correlations, using our database developed
in \cite{yonetoku10}. We examine this model bias by dividing
the samples into two data sets as gold and bronze according to
the quality of spectral observation.
In this paper, we assume, if signal-to-noise ratio is high enough, all of the spectrum of GRBs are well 
expressed by the Band function,

The structure of this letter is as follows. First we describe
our database of 109 GRBs with known redshift and well-determined
spectral parameters, 1-second peak luminosity, and isotropic energy
in section~\ref{sec:data}. We construct the \tsutsui, \yonetoku, and \amati
correlations with only gold data set in
section~\ref{sec:correlations}. Finally we will give
summary in section~\ref{sec:summary}.

\section{Data Description}
\label{sec:data}

In \citet{yonetoku10}, we constructed a database
selecting 109 GRBs from GCN Circular Archive \citep{gcn} and GRBlog
\citep{quimby03}. In this section we briefly describe our database.

Let us begin with $E_p$. In many cases, the prompt gamma-ray spectrum
is well fitted with the spectral model of the exponentially-connected
broken power-law function suggested by \citet{band93}.
This Band function has four parameters, the low-energy photon index
$\alpha$, the high-energy photon index $\beta$, the spectral break
energy $E_0$ and the normalization $A$. The peak energy ($E_{p}$), at which
the flux is maximum in the $\nu F_{\nu}$ spectrum, can be calculated
as $E_p = (2 + \alpha) E_0$.

However, for some GRBs, the photon index (mostly $\beta$) cannot be
determined due to the limited energy range of the detector and/or
the lack of the number photons \citep{pendleton97}. When the observation of
high-energy range is not enough, the spectrum is sometimes fitted
with the Cut-off power law (CPL) function. 
This function has three parameters, the low-energy photon index
$\alpha$, the spectral break energy $E_0$ and the normalization $A$.
In this case the peak energy can be derived as $E_p = (2 + \alpha) E_0$.
Note that even if, for a given GRB spectrum, the reduced chi square
value of this model is smaller than that of the Band function,
it is difficult to say whether this model reflects the intrinsic
property of the GRB or it is just due to the poor statistics
in the high-energy range. The reported values of $E_p$ for GRBs
which were poorly observed in the high-energy range are based
on either the Band function or CPL function, depending on the
observation team so that there exists the ambiguity in the definition of $E_p$
from the biginning. 

Let us move on to $E_{\rm iso}$, $L_p$ and $T_L$. In \citet{yonetoku10},
we calculated the bolometric energy and the peak luminosity
in the energy range 1-10,000~keV in the rest frame of each GRB
by extending the observed spectrum. 
Here, it should be noted that the integration
was performed assuming the Band function even for GRBs whose
spectra were not fitted by the Band function and the photon indices
were not reported. In these cases we assumed the typical values
$\alpha = -1$ and $\beta = -2.25$ to calculate 
the bolometric fluence ($S_{\rm bol}$) 
and the bolometric peak flux $F_{\rm p,bol}$. 
These values are suggested by BATSE observations
\citep{preece00} and also supported by Fermi observations of
GRB~080916C, 081024B, 090323 and 090428 up to possibly 100~GeV energy range.
Then the bolometric isotropic energy ($E_{\rm iso}$) and the 1-second
peak luminosity ($L_p$) can be simply calculated as
$
E_{\rm iso} = 4 \pi d_L^2 S_{\rm bol}/(1+z)~{\rm (erg)},
$
and 
$
L_p = 4 \pi d_L^2 F_{\rm p,bol}~{\rm (erg~s^{-1})}.
$
Here, $d_L$ is the luminosity distance calculated for the flat universe 
with the cosmological parameters of
$(\Omega_{\rm m}, \Omega_{\Lambda}) = (0.3, 0.7)$ and the Hubble
parameter of $H_0 = 70~{\rm km~s^{-1} Mpc^{-1}}$. Further we define
the luminosity time as the third parameter of GRB prompt emission as
$
T_L \equiv E_{\rm iso}/L_p.
$
The error of the luminosity time is estimated by using error
propagation equation. We can neglect the crossterm between 
$L_p$ and $E_{\rm iso}$ because of the independence of 
the \amati and \yonetoku relation shown in \citep{tsutsui09}. 

Thus, for GRB whose observed photon number is small, there are two possible
systematic effects. One comes from the fact that the peak energy
$E_p$ is determined by fitting the spectrum with either the Band
function or CPL function. As \citet{kaneko06} pointed out that
the CPL function tends to overestimate $E_p$ compared to
the Band function. This would induce a systematic error in
the correlations related to $E_p$. On the other hand, although
$L_p$ and $E_{\rm iso}$ are determined in a single straightforward way,
the photon indices are set to the typical values if the number of detected
photons is small. This would also cause a systematic error.

To estimate these systematic errors, we call a certain GRB belongs to the
gold data set if its  spectrum is well observed so that
it is  fitted by the Band function quite well  and all four parameters are
accurately determined. Other GRBs for which the  fixed $\alpha$ and/or $\beta$
are allocated to bronze data set. 
Here 3 of 109 GRBs in the database of \citet{yonetoku10}, do not have $E_{\rm iso}$
and then $T_L$ so they are  included in neither the gold nor the bronze
data sets.  As a result, the number of the gold and bronze data sets are 41 and 65 GRBs, respectively.

In the following sections, we construct the \amati, \yonetoku
and \tsutsui relations for the gold data set.
It is expected that the correlations obtained from the gold data set
would suffer from relatively small systematic errors so that
we could  study real intrinsic dispersions of the correlations.

\section{correlations}
\label{sec:correlations}

Here we derive the \tsutsui, \yonetoku, and \amati correlations from
the gold data set. 

First we assume the correlation among
$E_p$, $T_L$ and $L_p$ to be of the form,
$\log{L_p} \equiv A + B \log{(E_p/440 ~{\rm keV})} + C \log{(T_L /4.70~{\rm s})}$,
where we take the denominator of the second and  third terms 
as the average value of $E_p$ and $T_L$ in gold data set 
to minimize the statistical errors for these correlations.
For this correlation, we found
seven outliers (980425, 980613, 000131, 090328, 091003, 091020, 091127) shown by
blue color in top left of Fig.1.
The seven outliers  deviate from the best-fit relation at more than 3-$\sigma$ 
dispersion level. We will give some arguments on this point
in section~\ref{sec:summary}.
Using 34 gold data set of GRBs, we calculate the best fit function shown by the solid black line in top left of Fig. 1 
with red points of the gold data set
and 3 $\sigma$ errors of the \tsutsui relation by the yellow color region. 
The functional form of the best fit function is given by 
\begin{equation}
\label{eq:tsutsui}
L_p = (52.64\pm0.03) \times
     \left( \frac{E_p}{440 {\rm keV}} \right)^{1.70\pm0.07}
     \left( \frac{T_L}{4.70 {\rm sec}} \right)^{-0.40\pm0.06}.
\end{equation}
Here, we include not only errors in $L_p$ but also errors in 
$E_p$ and $T_L$ so that the chi-square function is defined as
$
\chi^2(A,B,C)
= \Sigma({\log{L_p^{\rm obs}}- A - B \log{(E_p/440 ~{\rm keV})} - C \log{(T_L /4.70~{\rm s})}})^2
  / (\sigma_{\rm meas}^2 + \sigma_{\rm int}^2)
$
where the first term of weighting factor is
$
\sigma_{\rm meas}^2
= (1+2C) \sigma^2_{\log{L_p}} + (B \sigma_{\log{E_p}})^2
 + (C \sigma_{\log{T_L}})^2
$. 
The factor 2C in  front of $\sigma^2_{\log{L_p}}$ comes
from the fact that the definition of $T_L$ includes $L_p$.
The reduced chi-square is unity with the intrinsic dispersion
$\sigma_{{\rm int}} = 0.13$. 
This correlation is consistent with our previous study \citet{tsutsui09}.

Similarly, we can obtain the best-fit function and errors of
the \yonetoku(top right of Fig.1) and \amati (bottom of Fig.1) correlations for the  same 34 gold GRBs,
\begin{equation}
\label{eq:yonetoku}
L_p = (52.63\pm0.05) \times 
     \left( \frac{E_p}{440~{\rm keV}} \right)^{1.76\pm0.10},
\end{equation}
with the intrinsic  dispersion $\sigma_{\rm int} = 0.22$, and 
\begin{equation}
\label{eq:amati}
E_{\rm iso} = (53.31\pm0.06) \times
     \left( \frac{E_p}{440~{\rm keV}} \right)^{1.68\pm0.13},
\end{equation}
with the intrinsic dispersion $\sigma_{\rm int} = 0.31$.
The best fit values of the \yonetoku and \amati correlations are consistent with previous studies but 
the intrinsic dispersions are tighter than those in \citet{yonetoku10} in which both gold and bronze
data are used.

The values of $\sigma_{\rm int}$ suggest that \tsutsui correlation is tightest among three correlations.
In Fig. 1, we show the gold data set (red points) with the best-fit function (solid line)
and 3-sigma dispersion region (dotted lines). We see that the \tsutsui correlation is much tighter than
the \yonetoku and \amati correlations by eye also. 
The blue points and green points indicate seven outliers of the \tsutsui correlation and 
bronze data set, respectively.
The bronze data set are systematically harder and/or dimmer than the gold data set.
This difference causes a large dispersion  in addition to the intrinsic dispersion 
of the correlations if we include the bronze data in the analysis.

\begin{figure*}
\rotatebox{0}{\includegraphics[width=80mm]{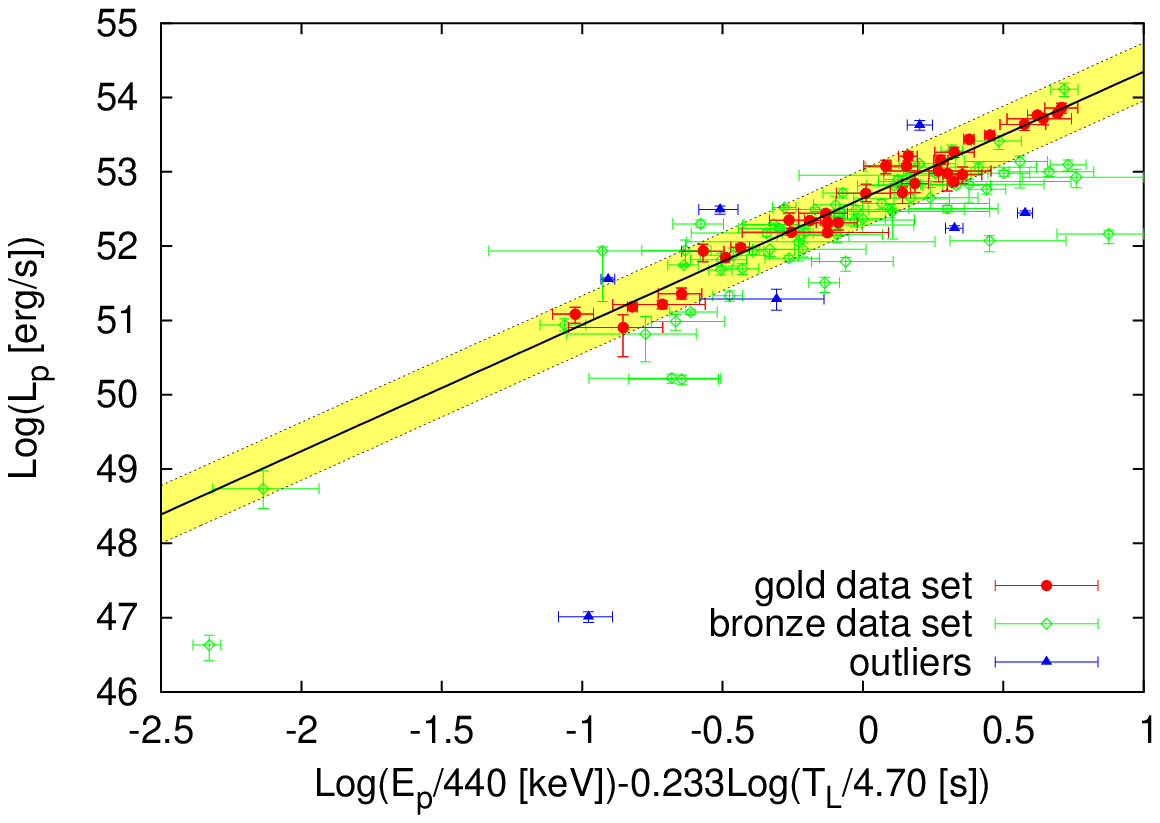}}
\rotatebox{0}{\includegraphics[width=80mm]{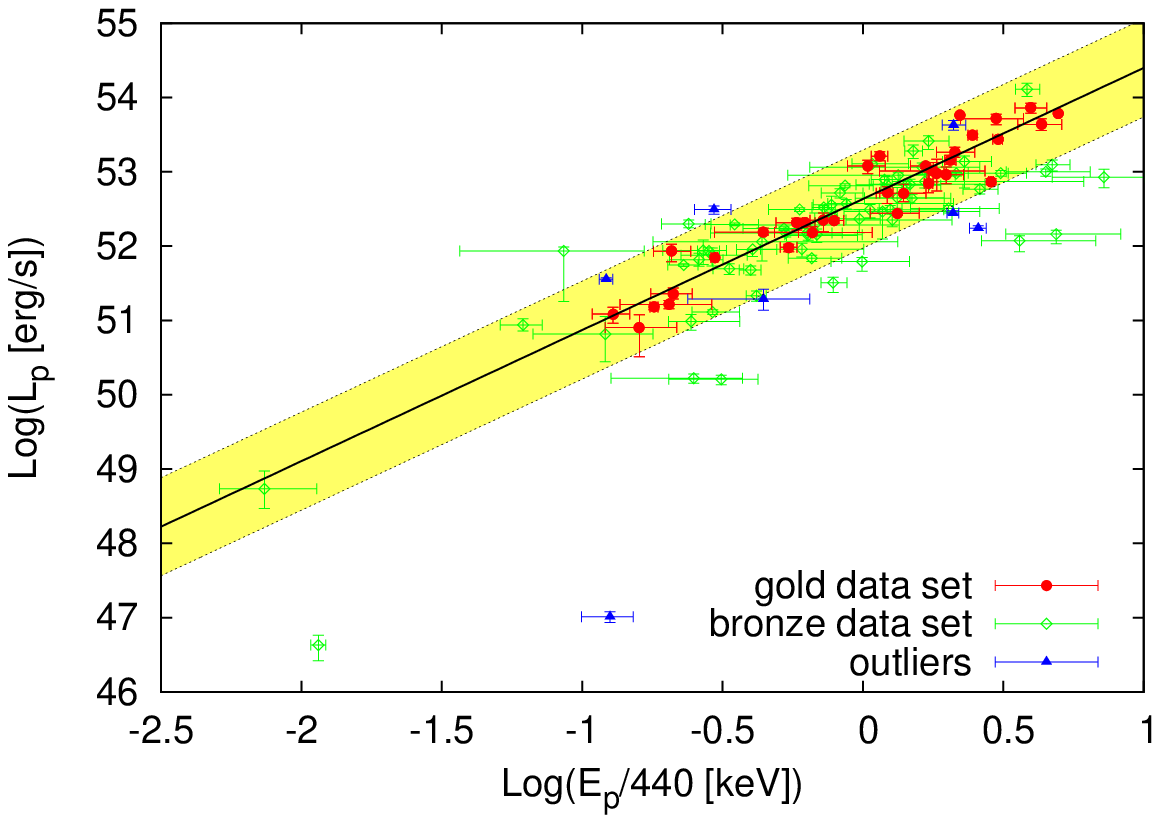}}
\rotatebox{0}{\includegraphics[width=80mm]{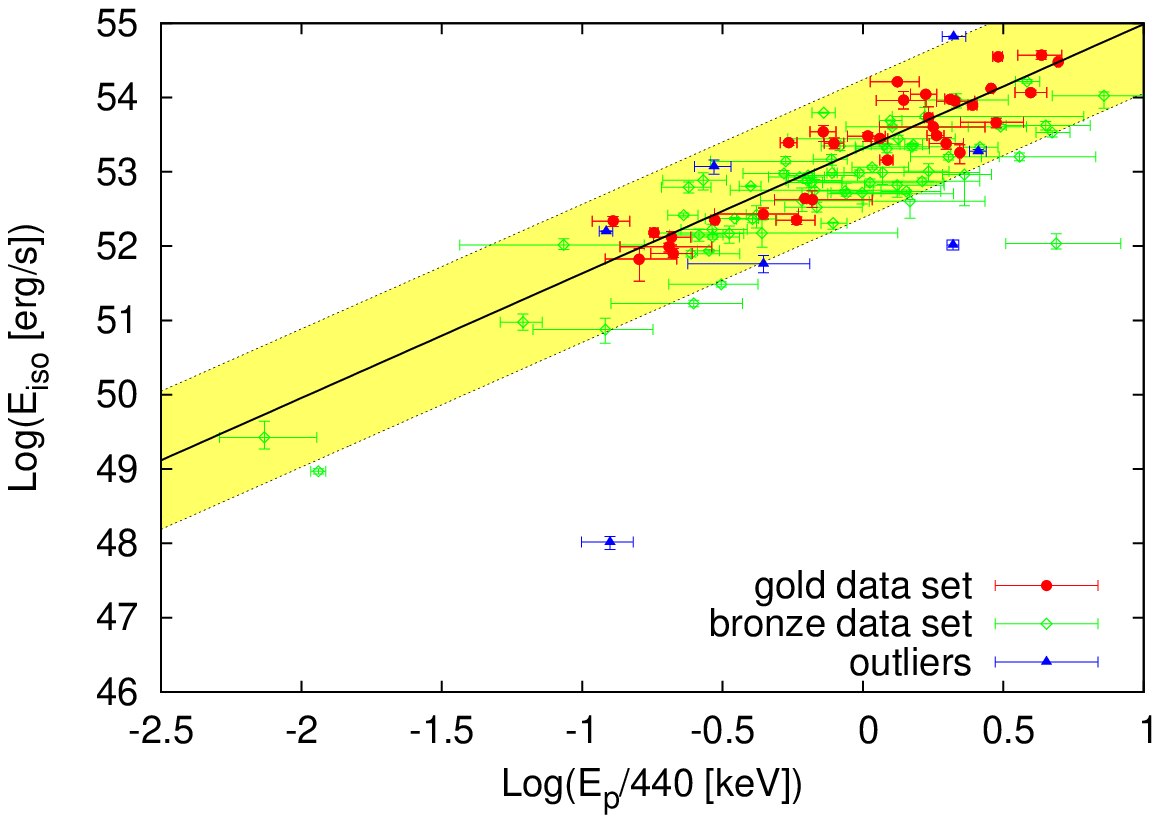}}
\caption{The \tsutsui relation (top left), the \yonetoku correlation
(top right) and the \amati (bottom) correlation with all data set. The solid line and dotted lines indicate
the best fit function and 3-$\sigma$ dispersion region in
Eq~(\ref{eq:tsutsui}), Eq~(\ref{eq:yonetoku}) and Eq~(\ref{eq:amati}). The bronze data
set seems to  be harder and/or dimmer than gold data set.
The CPL model or the Band function with fixed power-law index
cause this systematic difference.}
\label{fig:all}
\end{figure*}
\section{Summary \& Discussion}
\label{sec:summary}

In this paper, using database constructed by \citet{yonetoku10},
we examine the model bias, that is, Band or CPL, on  \tsutsui, \yonetoku and \amati correlations. 
We found that GRBs with the peak energies fitted by the CPL model 
are distributed in systematically harder and/or dimmer side of the \tsutsui, \yonetoku and \amati correlations 
than the ones by the Band function. 
There might be two interpretations about this result. 
The first is that these correlations  have much larger intrinsic dispersion than that of 
observed one. If we had the more sensitive detector and could observe dimmer GRBs, 
the dispersion of the relations would become larger\citep{butler,sn09}.
Another is that the use of  the CPL model to estimate the peak energies causes 
this systematic difference.
As simulated by \citet{kaneko06},  the Band function spectrum
is well fitted by the CPL model if  the detector does not have enough sensitivity 
to observe the  high-energy photons. However the peak energies fitted by the 
CPL models are always higher than that of the simulated Band function spectrum \citep[see table.~3 in ][]{kaneko06}.
Thus, it seems to be natural to conclude that the latter is more acceptable.
In short, using only the peak energies  determined by the Band function, 
we would get tighter correlations. 
If we could have much more GRBs by which we can uniformly analyze the data with the 
Band function, GRBs would be more powerful tool to constrain cosmological parameters.

We found seven outliers in our gold data set. We  classify 
these outliers in two classes as  
\begin{enumerate}
\item [](dimmer and/or harder)  980425, 980613, 090328, 091003
\item [](brighter and/or softer)  000131, 091020, 091127
\end{enumerate}
Although we do not  know how and why  these outliers are  different from ordinary GRBs
except for the distribution in the \tsutsui space, 
the effect of eliminating these GRBs is obvious, that is, the correlation becomes tighter.
To find the characteristics which distinguish these outliers from ordinary GRBs
is urgent. We here point out  possible origins of these outliers.
Let us assume that if we observe the jet nearly on axis ,  \tsutsui correlation would be very tight.
However if we observe the jet with a certain  viewing angle, we might have  some 
dispersions on the observed \tsutsui correlation.
In other words, we might not  avoid some dispersion in the \tsutsui correlation from  viewing angle, 
especially when the observer locates near the edge of  the jet of GRBs. 
If we will know how to distinguish  these outliers from the ordinary gold GRBs,
the \tsutsui correlation might be much tighter and very useful in determining the nature of dark energy 
in redshift larger than $\sim$3.
We should note that  even in the Period-Luminosity relation of Cepheid variable  
there are $\sim$ 10\% outliers \citep{Riess09} so it is not surprising that there are $\sim$ 20\% 
outliers in the \tsutsui relation. 

\citet{butler}, using the Bayesian approach to estimate $E_p$, 
indicated that dim events close to the detector sensitivity would 
make large scatter on the $E_p$--$E_{\rm iso}$ and $E_p$--$L_p$ 
relations and that there is a significant threshold effect.
Thus, they conclude that the \amati correlation have larger 
intrinsic dispersion than observed if we do not suffer from a threshold effect.
Recently, \citet{sn09} argue that using hardness ratio instead of $E_{p}$ they 
also find that $E_{p}^{obs}$-Fluence correlation become more wider if we will 
be able to determine $E_{p}$ of dimmer events. 
However, there is a possible bias by using different method to
estimate $E_p$. 
Even the difference of $E_{p}$ between the Band and CPL models causes the 
systematic errors so that using the other method to estimate $E_{p}$ might cause the additional 
systematic errors.
The smaller the intrinsic dispersion of the relation is, the more the correlation suffers from these
systematic effects. This might be why the $E_{p}$--$T_{0.45}$--$L_{p}$ relation is not confirmed 
by later studies \citep{firmani06, rossi08,collazzi08}.

Kaneko et al. (2006) suggested the $E_{p}$ value of CPL
function becomes systematically higher than 
the one of the Band function. 
If the 65 bronze data previously analyzed by
the CPL function are reconsidered by the Band function
with the fixed $\beta$ as an average value of $-2.25$
(Preece et al. 2000), they might show the distribution
around the best fit line of each correlation 
estimated with 41 gold data set.
They have a good potential to become a "silver" data set.
To do so, we need help from each instrument team, 
and this is a future work.

Finally, we note that there would be many reasons which cause systematic errors on the correlation in addition 
to intrinsic property of GRBs. These systematic errors must  be carefully estimated and removed from the 
correlation analysis one by one. If we will finish it, GRBs become more powerful and unique 
standard candles to investigate the nature of the dark energy at high redshift larger than $\sim$ 3.

\section*{Acknowledgments}

This work is supported in part by the Grant-in-Aid from the 
Ministry of Education, Culture, Sports, Science and Technology
(MEXT) of Japan, No.19540283, No.19047004(TN),
No.18684007 (DY) and No.21840028(KT), and by the Grant-in-Aid
for the global COE program {\it The Next Generation of Physics,
Spun from Universality and Emergence} at Kyoto University and
"Quest for Fundamental Principles in the Universe: from Particles
to the Solar System and the Cosmos" at Nagoya University
from MEXT of Japan. RT is supported by a Grant-in-Aid for
the Japan Society for the Promotion of Science (JSPS) Fellows
and is a research fellow of JSPS.

\end{document}